\documentclass[journal=inoraj,manuscript=article]{achemso}
\usepackage{graphicx,bm}
\usepackage{dcolumn}
\usepackage{bm}
\usepackage{upgreek}
\usepackage{times}
\usepackage[caption=false]{subfig}
\usepackage{mathrsfs}
\usepackage{amsmath} 
\usepackage[version=3]{mhchem} 


\title{Short-range ordering in the Ni-Mn-Si based Laves phase Mn(Ni$_{0.6}$Si$_{0.4})_2$}

\author{S. J. Ahmed}
\affiliation{Department of Materials Science and Engineering, McMaster University, Hamilton, Ontario.}

\author{M. Niewczas}
\affiliation{Department of Materials Science and Engineering, McMaster University, Hamilton, Ontario.}
\alsoaffiliation{Brockhouse Institute of Materials Research, McMaster University, Hamilton, Ontario.}
\email{niewczas@mcmaster.ca}

\date{\today}

\begin{document}
\begin{abstract}
We present the structural ordering and the associated physical behavior of a Ni-Mn-Si Laves phase, Mn(Ni$_{0.6}$Si$_{0.4})_2$. The high-resolution transmission electron microscopy and electron energy loss spectroscopy analysis were performed to resolve a distinct atomic ordering of the system. The study determined the origin of the short-range ordering to be the unique arrangement between Ni and Si atoms. The study also presents the atomic resolution mapping of the Si atoms which has never been reported by any previous studies. With further electrical conductivity measurement, we find one of the consequences of the unique ordering reflected in a semiconducting like temperature dependence of the compound.
 \end{abstract}

\maketitle

\section{Introduction}

 With more than 1400 reports of stable binary and ternary compound, Laves phase (also known as Friauf-Laves phases~\cite{Friauf1927a,Friauf1927,Laves1939}) contains the largest group of the intermetallic family~\cite{Stein2004}. From different types of Laves phase, one of the most technologically important groups are the silicon containing ternary compounds in which the addition of a third element (Si) in a non-Laves transition metal compound stabilizes a C14 type Laves phase structure~\cite{Bardos1961,Bardos1963,Savitskii1965,Markiv1966,Mittal1978}. Such modification has been a topic of great scientific interest as the resulting transition metal-silicon compound demonstrate superior magnetic, electronic, thermal and mechanical properties\cite{Mittal1978,Stein2004,Yan2007,Bhowmik2011,yan2013phase}. The stabilization is believed to be due to a reduction of effective electron concentration by absorption or localization of the transition metal electrons by silicon~\cite{Bardos1961,W.Hume-Rothery1965,Mittal1978}.\\


There has been very limited number of studies on the structural and physical properties of Laves phase containing Ni, Mn and Si~\cite{Gladyshevskii1956,Y.B.Kuzma1964,yan2013phase}. Phase equilibria of the Ni-Mn-Si system suggested the phase to be stable within a wider homogeneity range of 26-30 atomic percent of Si, 31-35 atomic percent of Mn and balance Ni~\cite{Bardos1966,Hu2011}. A comprehensive study of structural and magnetic properties on Ni-Mn-Si Laves phase was performed by \citet{yan2013phase} where a Mn(Ni$_{0.625}$Si$_{0.375})_2$ composition was investigated. The system was found to be an antiferromagnet with a N\'{e}el temperature of 630 K, and the authors proposed a possible magnetic configuration for it. The study reported the preferential site occupations of Ni and Si in 2a and 6h where the two sites were found to facilitate the atoms in unconventional proportion. Such a distinct atomic arrangement is expected to give rise for some special atomic ordering in the system. The presence of short-range ordering in Laves phase systems is a very well known phenomenon that has been observed as diffused scattering in the diffraction intensity in many other studies~\cite{Komura1962,Komura1980,Skripov1999,Irodova2000,Kelton2003,Skripov2004}. However, to the best of our knowledge, there has been no attempt to make a structural reconstruction of such ordering.\\

In this work, we conducted a study on the atomic ordering in the single crystal of the Laves phase with Mn(Ni$_{0.6}$Si$_{0.4})_2$ composition. High resolution transmission electron microscopy (HRTEM) and electron energy loss spectroscopy analysis (EELS) analysis was performed to determine the ordering of atoms within the crystal. Further transport property measurement was carried out which was found to be linked with the resolved atomic ordering.



\section{Experimental details}\label{sec:Experimental details}

Single crystal of Mn(Ni$_{0.6}$Si$_{0.4})_2$ Laves phase was grown by RF heating Czochralski method in an argon atmosphere. Pure elements of Ni (99.95\%), Mn(99.98\%) and Si(99.999\%) were melted in an alumina crucible and the single crystal was pulled using a tungsten wire seed with a constant pulling rate of 0.5 mm/min along with a 30 rpm rotation. With the Laue diffraction, the crystal was found to be grown along $[0001]$.\\

The single-crystal diffraction (SCD) was done on a tiny crystal piece about 100 $\mu$m diameter and 50 $\mu$m thickness, using the Bruker Apex II Diffractometer with Mo K$\alpha_{1}$ radiation and Apex II CCD Detector. The MAX3D~\cite{Britten2007} software was used to visualize the reciprocal space that confirmed the crystallinity of the material. The crystal structure was solved and refined using the SHELXS and SHELXL~\cite{Sheldrick1997} software packages.\\

High-resolution scanning transmission electron microscopy (STEM) and electron energy loss spectroscopy (EELS) studies were conducted on a (0001) cross section of Mn(Ni$_{0.6}$Si$_{0.4})_2$ single crystal sample using an aberration-corrected (probe-forming lens) FEI Titan 80-300 LB Cube microscope operating at 200 kV and equipped with a Gatan GIF Quantum electron energy-loss spectrometer. High angle annular dark-field (HAADF) images have been acquired using  K2 summit direct electron detection camera~\cite{Subramaniam2016}, with a convergence angle of 19 mrad and a collection angle in the range of 64 - 200 mrad. Electron transparent foils suitable for HAADF-STEM studies were prepared by cutting 400 $\mu$m thick (0001) sections of the crystal with electro-discharge machine. The slices were mechanically thinned down to the thickness of about 100 $\mu$m, using set of grinding papers and Struers LaboSystem. 3 mm diameter disks were spark cut form the slices and used for further preparation of the foils. The thinning of the discs was conducted with dimpler, Gatan 656, down to about 10 $\mu$m thickness using the diamond abrasive wheel lubricated by diamond paste. The final thinning of the discs was conducted using ion beam polishing system Fischione 1010, at 78 K and 6 kV for $\sim$6 hours. \\

The electrical conductivity was measured on a single crystal sample with gauge dimensions 10 mm$\times$4 mm$\times$2 mm, between 1.8 K and 300 K, with the Keithley 2182A nanovoltmeter and 6221 current source, attached to the Quantum Design PPMS system. A four-point method of measuring potential drop across the sample was used to determine the sample resistance. The sample was mounted on a platform with a spring-loaded, point-contact potential and current leads. The potential drop was measured in Delta mode as the average of 100 current reversals.


\section{Results}\label{sec:Results}

\subsection{Single crystal diffraction}\label{sec:Single crystal diffraction}

The refined structural and the atomic parameters of Mn(Ni$_{0.6}$Si$_{0.4})_2$ compound are listed in table~\ref{Table:MnNi1.2Si0.8_SC_XRD_298K} and table~\ref{Table:MnNi1.2Si0.8_SC_XRD_atomic298K}, respectively. The parameters are found to be in good agreement with the previously published results on a slightly different composition MnNi$_{1.25}$Si$_{0.75}$ Laves phase~\cite{yan2013phase}. The crystal structure was solved in a MgZn$_{2}$-type structure (C14 Laves phase) in the $P 63/mmc$ spacegroup. Best refinement was obtained with the atomic parameters such that the Mn atoms occupy the $4f$ crystallographic site while Ni and Si are located on the $2a$ site with 0.471/0.529 fractional occupancy and $6h$ position with occupancies of 0.66/0.33, respectively. The occupancy parameters obtained are consistent with the crystal chemistry of the MgZn$_{2}$-type C14 Laves phase. Note, in these compounds atoms with the largest radius are located at the $4f$ position (Mn in the present case with the atomic radius of $1.61\,{\AA}$). Also, in the ternary Laves system, the $2a$ and $6h$ occupancies were found to be distributed in a manner that the atoms with the highest concentration prefer to be in the $6h$ site which is Ni in the present case~\cite{Kerkau2012}.


 \begin{table}[htp]
    \caption{Crystallographic data and refinement parameters for the Mn(Ni$_{0.6}$Si$_{0.4})_2$ single crystal (Mo K$_\alpha$ radiation, 298K)}\label{Table:MnNi1.2Si0.8_SC_XRD_298K}
     \centering
        \begin{tabular}{l l l l l }
        \hline
        \hline

              &   &  \\
            \hline
                & Refined composition &Mn$_{4}$ Ni$_{4.96}$Si$_{3.04}$ \\
                &Space group& $P 63/mmc$ \\
                &Lattice constant $({\AA})$& $a=4.7639(1)$, $b=4.7639(1)$ $c=7.4967(2)$ \\
                &Volume $({\AA}^{3})$& $147.342(7)$ \\
                &$\rho_{calc} (g/cm^3))$& $6.666$ \\
                &Z & $4$ \\
                &$2\theta$ range & $9.882 - 90.434$ \\
                &Index ranges& $-8\leq h\leq9$,$-9\leq k\leq8$$-14\leq l\leq-11$ \\
                &Reflections collected& $2373$ \\
                &Independent reflections& $271$$[R_{int}=0.0389,\,R_{sigma=0.0232}]$ \\
                &Data/restraints/parameters& $271/0/13$ \\
                &Goodness-of-fit on $|F|^2$& $1.208$ \\
                &Largest diff. peak/hole (e/${\AA}^{3}$)& $1.23/-1.18$ \\
                &R indices $[I>=2\sigma(I)]$& R$_1$=0.0222 wR$_2=$0.05\\

                \hline
                \hline
        \end{tabular}
\end{table}

 \begin{table}[htp]
    \caption{Fractional atomic coordinates, occupancy and isotropic displacement parameters for Mn(Ni$_{0.6}$Si$_{0.4})_2$, obtained from the refinement of Mn(Ni$_{0.6}$Si$_{0.4})_2$ single crystal diffraction data.}\label{Table:MnNi1.2Si0.8_SC_XRD_atomic298K}
     \centering
        \begin{tabular}{l l l l l l l l }
        \hline
        \hline
        &Atom& x & y&z& Site&Occupancy&U$_{eq}\,({\AA}^{2})$\\
        \hline
        &Mn1&0.3333&0.6667&0.0648(4)&4f&1&0.00720(11)\\
        &Ni1&0&0&0&2a&0.471(5)&0.01208(16)\\
        &Si1&0&0&0&2a&0.529(5)&0.01208(16)\\
        &Ni2&0.82873(3)&0.65746(6)&0.25&6h&0.67(5)&0.00704(9)\\
        &Si2&0&0&0.5&6h&0.33(5)&0.00704(9)\\
                \hline
        \end{tabular}
\end{table}

Analysis of the three-dimensional reciprocal space using MAX3D~\cite{Britten2007} revealed triangular-shape diffuse scattering pattern along $a^*b^*$ planes, which is an indication of short-range ordering (SRO) of atoms within the $ab$ planes. In Mn(Ni$_{0.6}$Si$_{0.4})_2$, Mn atoms occupy the 4f site and are expected to order periodically. Therefore, it can be assumed that Ni and Si atoms fill the remaining  2a and 6h positions in a unique short-range order that gives rise to the observed diffused diffraction pattern.


\begin{figure}[htp]
\centering
   \includegraphics[scale=0.5]{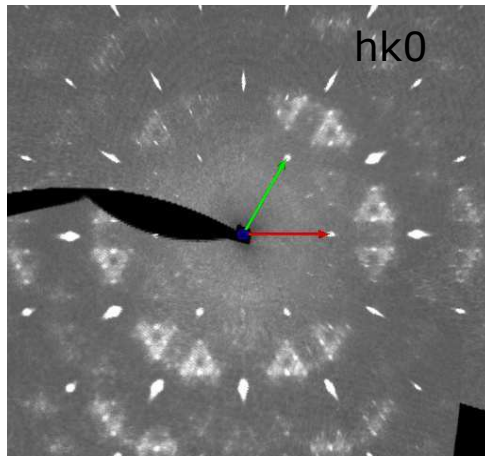}\\
   \caption{Diffraction pattern of Mn(Ni$_{0.6}$Si$_{0.4})_2$ along (hk0) plane showing triangular-shape diffused scattering patten.}
   \label{fig:triangles_hk0}
\end{figure}

\subsection{Transmission electron microscopy analysis (HRTEM and EELS)}
\label{sec:Transmission electron microscopy analysis}

\begin{figure}[htp]
\centering
   \includegraphics[scale=0.6]{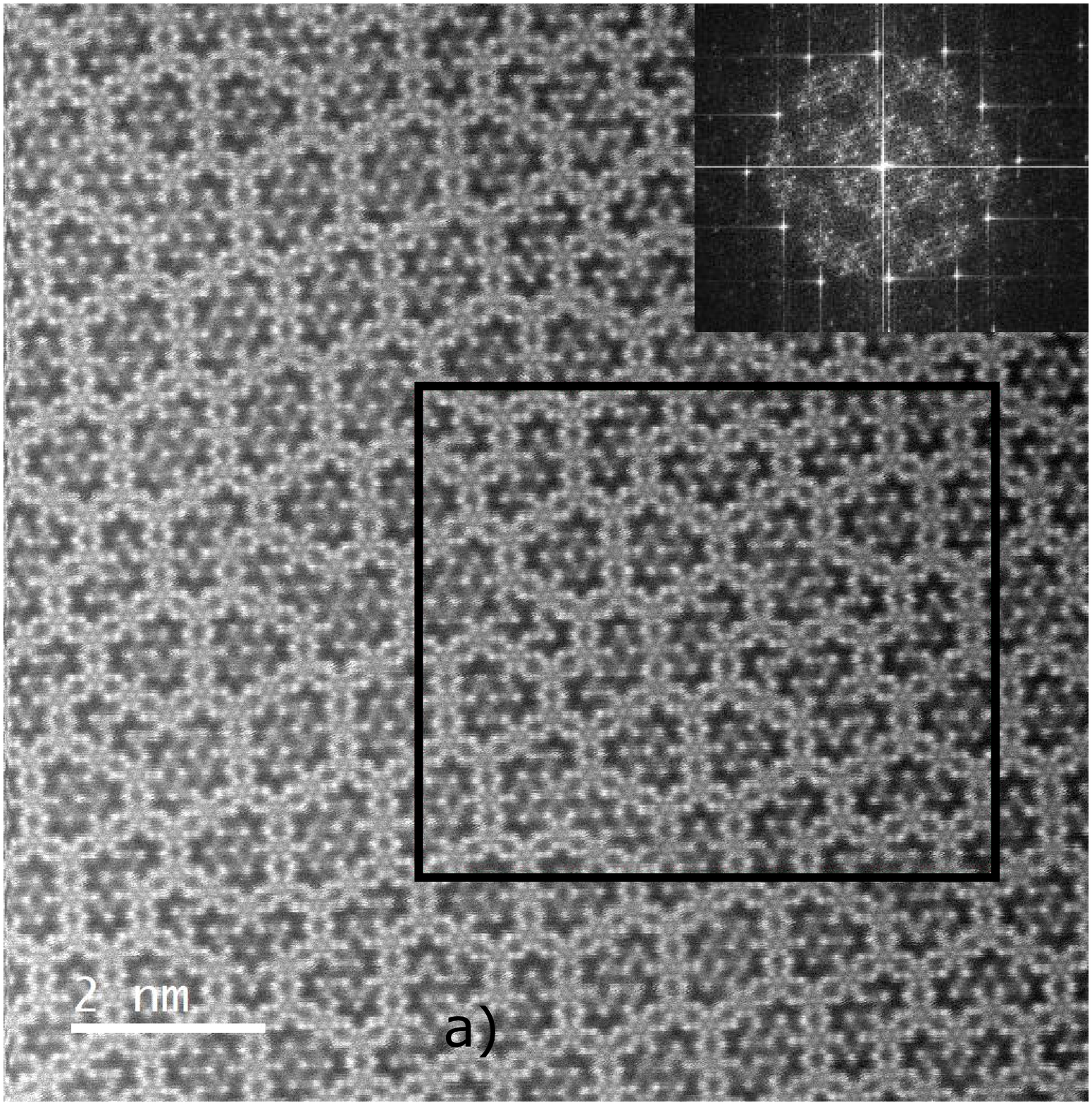}\\
   \caption{HAADF-STEM image of (0001) section of Mn(Ni$_{0.6}$Si$_{0.4})_2$ compound. Inset shows diffused scattering pattern obtained by the Fast Fourier Transform (FFT) of the image. Box marks the area of EELS analysis.}
   \label{fig:STEM_diffuse}
\end{figure}

High-resolution HAADF-STEM image (Fig.~\ref{fig:STEM_diffuse}) of the (0001) section of Mn(Ni$_{0.6}$Si$_{0.4})_2$ shows the presence of a distinctive short-range ordering pattern for the compound. It can be noticed that the atomic arrangement within the structure are accommodated in a manner that several non-periodic units have been formed. These small units of different sizes are arranged randomly in a configuration that does not permit to assign a periodic unit cell. The Fast Fourier Transform (FFT) of the image (inset in Fig.~\ref{fig:STEM_diffuse}) shows a diffused scattering pattern similar to that observed in the single crystal diffraction. The less bright elements can be identified as Si because of their low atomic number of Z=14. Ni and Mn, on the other hand, possess similar atomic numbers of Z=25 and Z=28, respectively. Therefore, the brighter Ni and Mn could not be distinguished in the HAADF-STEM image (Fig.~\ref{fig:STEM_diffuse}).\\
%

\begin{figure}[htp]
\centering
   \includegraphics[scale=0.8]{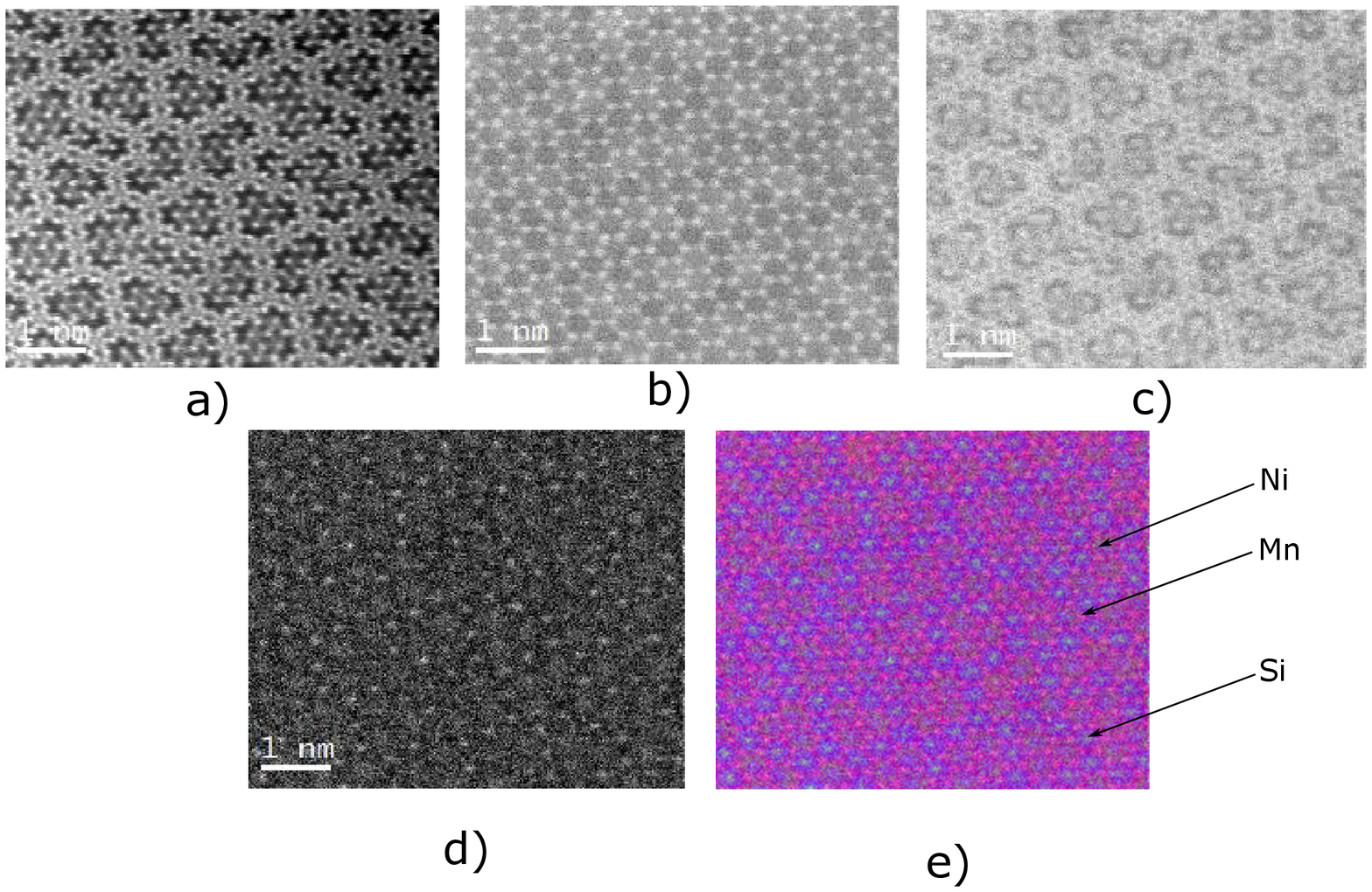}\\
   \caption{EELS atomic resolution mapping on the (0001) section of Mn(Ni$_{0.6}$Si$_{0.4})_2$ single crystal. a) HAADF image of the area of interest acquired in parallel with the EELS mapping. b) Mn L atomic resolution chemical map showing an ordered arrangement of Mn atoms. c) Ni atomic resolution map indicating a distinct non periodic arrangement of the atoms d) 2a Si atomic resolution map capturing the atoms to be located in spaces left by 2a. e) a composite image showing the unique short-range ordering in terms of 2a Ni and Si atoms and the Mn atoms located in 4f. Here the Ni atoms are colored in grey, Si in blue and Mn in pink.}
   \label{fig:STEM_EELS}
\end{figure}



 To identify the distribution of all elements, EELS maps were obtained in the boxed area of the Fig.~\ref{fig:STEM_diffuse} which are are shown in Figs.~\ref{fig:STEM_EELS}a-e. The Mn L signal (Fig.~\ref{fig:STEM_EELS}b) shows a homogenous distribution of the Mn atoms. The effective Mn-Mn distance was determined on the order of $\sim$2.6 (1)$\AA$ which is compatible with the horizontal distance between two nearest neighboring Mn atoms located in 4f sites. Consequently, it can be confirmed that not only the Mn atoms from the top plane (0001) but the atoms that lie on below are also captured in the elemental mapping. Note that, the Mn-Mn distance for the topmost plane is 4.76 (1) $\AA$ obtained from SCD refinement data.  The mapping of Ni atoms~\ref{fig:STEM_EELS}c shows an unique arrangement with the tendency to form a network of three or four or five neighboring atoms. Within the network, the average Ni-Ni nearest neighbour distance of $\sim$4.7$\AA$ is similar to the lattice constant of 4.7639(1)$\AA$ confirming that the Ni atoms captured in the mapping are located in the 2a positions. Figure~\ref{fig:STEM_EELS}d shows the elemental maps for the Si atoms which shows that rest of the $2a$ positions are occupied by the Si atoms. Si K signal is not usually detected in the EELS mapping. However, the elemental mapping for Si observed in the present work is due to the use of the powerful Gatan K2 Summit detector~\cite{Subramaniam2016} that provides unmatched contrast and resolution for all types of molecules. The combined image is shown in Fig.~\ref{fig:STEM_EELS}e  where the Ni atoms are colored in grey, Si in blue and Mn in pink. It can be seen that Ni and Si atoms occupy the 2a positions in a unique manner with a Si 2a chain enveloping the network of Ni 2a sites. The 6h atoms were not observed in EELS, which is due to their close horizontal spacing of $\sim$1$\AA$ being beyond the resolution of the EELS mapping at this magnification.\\

\begin{figure}[htp]
\centering
   \includegraphics[scale=0.5]{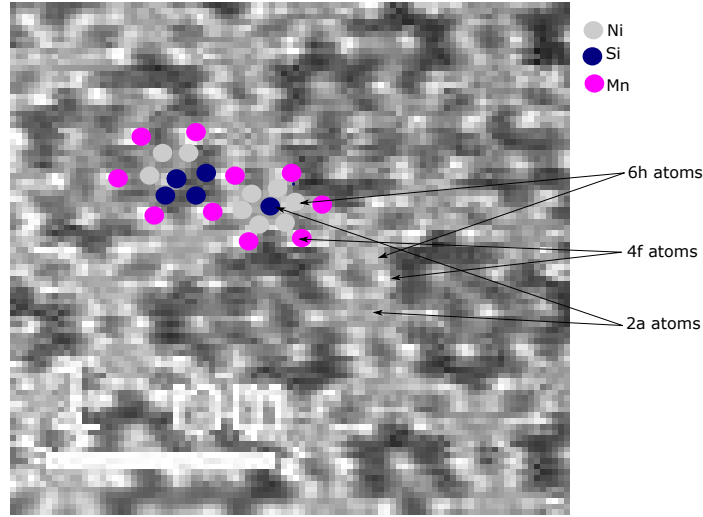}\\
   \caption{a) A magnified section of Mn(Ni$_{0.6}$Si$_{0.4})_2$ HRTEM image identifying the small tiny dots between the Mn formed hexagons as the Ni 6h atoms. The low scattering Si atoms appears as voids in the 6h positions.}
   \label{fig:STEM_ZOOM}
\end{figure}

However, a zoom-in HAADF image from the analyzed area in Fig.~\ref{fig:STEM_ZOOM} shows the presence of the small atoms within the hexagonal unit of Mn that can be identified as the atoms located in the 6h positions. It can be seen that the structural unit that contains Si atom in the 2a position (less bright hexagon center), have these small visible atoms in all the 6h position, which can be identified as Ni atoms. Weak resolution of 6h Ni atoms be explained by the close spacing of the atoms in the site. On the other hand, the structural unit with Ni in the 2a position contains tiny atoms and empty spaces. These can be thought to comprise of Ni (tiny atom) and Si (empty space). The 6h Si being invisible is not surprising as these low scattering atoms are lying on a lower planes with a smilar horizontal spacing as Ni in that position.


\subsection{Electrical conductivity measurement}
\label{sec:Electrical conductivity measurement}

\begin{figure}[htp]
\centering
   \includegraphics[scale=0.5]{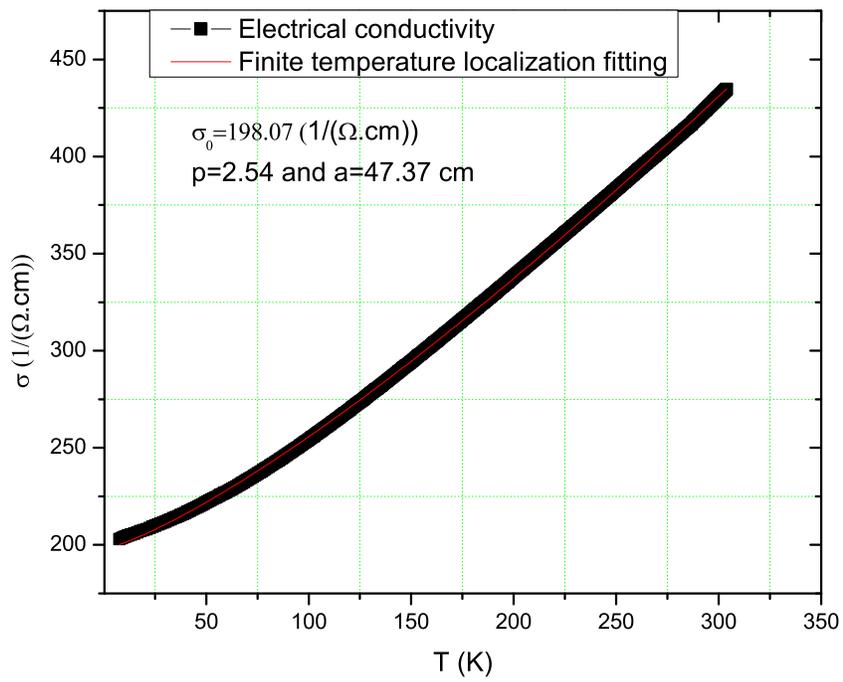}\\
   \caption{Temperature dependence of electrical conductivity of Mn(Ni$_{0.6}$Si$_{0.4})_2$ from 2 K to 300 K. Data was fitted to the model describing the weakening of quantum interference due to random fluctuations.}
   \label{fig:Weak_localization_fitting}
\end{figure}


Figure ~\ref{fig:Weak_localization_fitting} shows variation of the electrical conductivity with the temperature. The conductivity increases as the temperature increases which is a behaviour usually observed in semiconducting systems. A similar behavior of the electrical conductivity for Ni-Mn-Si Laves phase was reported by ~\citet{yan2013phase}. The authors linked this behaviour with the localization of conduction electron due to the disorder caused by Ni and Si occupancy. The present data was fitted to the same three-dimensional temperature dependent electrical conductivity equation for localization~\cite{Lee1985},

 \begin{equation}\label{Eq:3Dlocalized_conductivity}
\sigma_{3D} (T)=\sigma_{0}+\frac{e^{2}}{\hbar \pi^{2}}\, \frac{1}{a}T^{\frac{p}{2}}.
 \end{equation}

The fitting yielded the temperature independent conductivity, $\sigma_{0}$=198.07 $\frac{1}{\Omega .cm}$, scattering mechanism index, $p$=2.54 and a measure of length scale, $a$=47.37 cm. $p$ and $a$ are related to the length scale of the quantum inference, $L_{th}=aT^{\frac{p}{2}}$. $L_{th}$ is a measure of the weakening of localization by the random fluctuation due to finite temperature. The conductivity data thus provide evidence of localization in the system. The $p$=2.54 is an indication of electron-phonon interaction in the system to be a dominant mechanism responsible for limiting the quantum interference~\cite{Lee1985}.


\section{Discussion}\label{sec:Discussion}

\begin{figure}[htp]
\centering
   \includegraphics[scale=0.7]{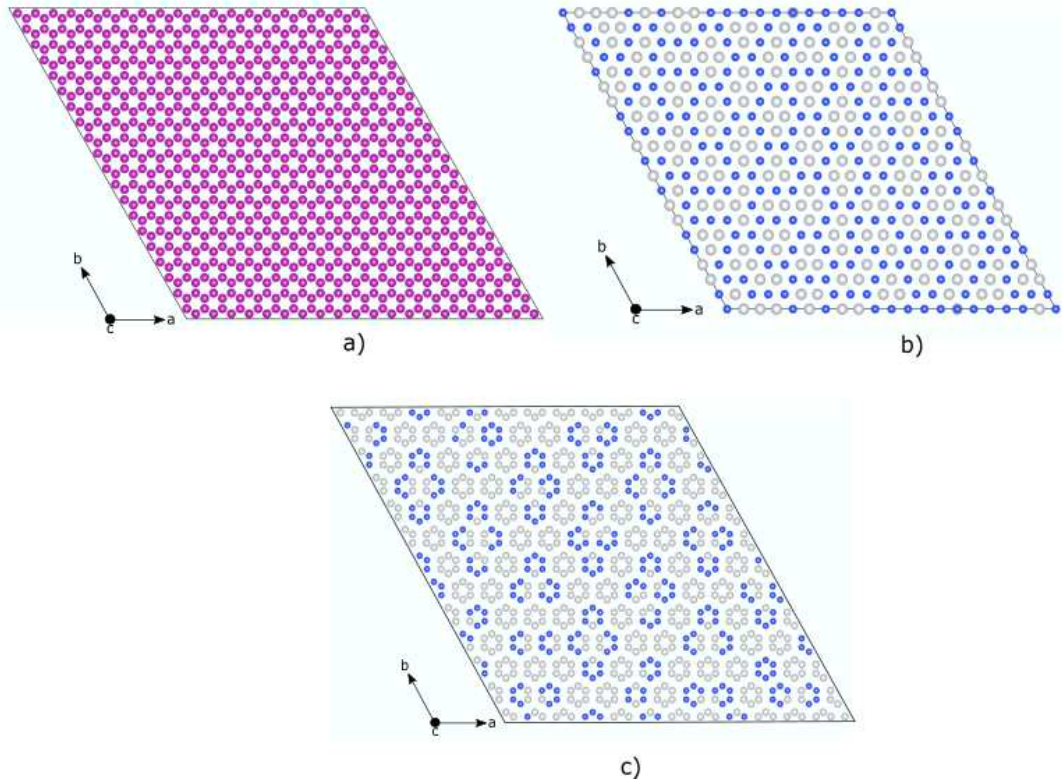}\\
   \caption{A prototype structure of Mn(Ni$_{0.6}$Si$_{0.4})_2$ generated based on the EELS analysis showing the distribution of  a) 4f Mn atoms to form a hexagonal pattern b) Ni and Si in 2a position in a unique manner where Si 2a forms a network containing the Ni 2a atoms inside and c) Ni and Si atoms in 6h.}
   \label{fig:simulated_Laves_structure}
\end{figure}

Based on the analysis of the STEM and EELS observations (Sec.~\ref{sec:Transmission electron microscopy analysis}), a representation of SRO structure of Mn(Ni$_{0.6}$Si$_{0.4})_2$ is shown in Fig.s~\ref{fig:simulated_Laves_structure} (a-c). The Mn atoms order periodically with full occupation of the 4f site (Fig.~\ref{fig:simulated_Laves_structure} (a)). The unique SRO in the crystal is mainly attributed to the arrangement of Ni and Si in 2a sites that gives rise to several non-periodic units (Fig.~\ref{fig:simulated_Laves_structure} (b)). These units consist of several small blocks which have Ni or Si 2a atoms as their centers. The 6h Ni and Si atoms are arranged depending on which block they belong to (Fig.~\ref{fig:simulated_Laves_structure} (c)). The Si centered blocks only contain Ni atoms in the 6h positions. While the Ni centered blocks contain the Ni and Si atoms arranged randomly in the 6h. Interestingly, it was observed that each of the non-periodic units have a composition that is somewhat similar to the actual stoichiometry of Mn(Ni$_{0.6}$Si$_{0.4})_2$. The analysis also points out that 2a sites are predominantly occupied by the Si atoms while the 6h sites by the Ni atoms. The observation is consistent with the occupancy obtained from single crystal refinement. The occupancy of the Ni and Si in the Wyckoff position 2a and 6h can be explained in terms of the explanation by Faller and Skolnick ~\cite{Faller1963} that the 2a sites are mostly occupied by the lowest concentration atom if a heteronuclear type bonds are preferred by the occupants of 2a and 6h. Amongst the Ni and Si in the ternary C14 Mn(Ni$_{0.6}$Si$_{0.4})_2$, the Si atoms are present in the lowest concentration. Therefore, the occupation of majority 2a sites by Si that is reflected in single crystal occupancy in Table~\ref{Table:MnNi1.2Si0.8_SC_XRD_atomic298K} and, in Fig.~\ref{fig:simulated_Laves_structure} (b) can be justified. It is also evident that the system prefers to form heteronuclear Ni-Si bonds more than the homonuclear Ni-Ni or Si-Si bond.

The EELS analysis in (Sec.~\ref{sec:Transmission electron microscopy analysis}) captured the atomic resolution mapping of Si in Fig. \ref{fig:STEM_EELS}d which is a notable observation. To the best of our knowledge, no other previous studies were able to show such accurate mapping of Si atoms.

The electrical conductivity behavior observed in Fig.~\ref{fig:Weak_localization_fitting} can be considered as a consequence of the unique ordering of the Ni and Si atoms in Mn(Ni$_{0.6}$Si$_{0.4})_2$. Despite a small range periodicity (Fig~\ref{fig:simulated_Laves_structure} b and c), the absence of the long-range order is strong enough to destroy the periodic potential which is reflected in the electrical conductivity measurement.

\section{Conclusions}

We resolved the short-range atomic ordering and characterized physical properties of the Ni-Mn-Si based Laves phase with the composition Mn(Ni$_{0.6}$Si$_{0.4})_2$. The atomic ordering determined with the HRTEM and EELS revealed a short-range atomic arrangement between Ni and Si atoms producing small blocks of unit cells that are non-periodic but have the Mn(Ni$_{0.6}$Si$_{0.4})_2$ stoichiometry. The EELS analysis in this work was also able to produce the atomic resolution mapping of Si which has not been detected before by any other research. During further physical property characterization, the study also reported a semiconductor like temperature dependence of the electrical conductivity for the metallic Mn(Ni$_{0.6}$Si$_{0.4})_2$ which was associated with short-range ordering that caused localization of conducting electrons.

\begin{acknowledgement}
Financial support of Natural Sciences and Engineering Research Council of Canada under the NSERC grant: "Artificially Structured Multiferroic Composites based on the Heusler alloys" is gratefully acknowledged.
\end{acknowledgement}

\providecommand{\latin}[1]{#1}
\makeatletter
\providecommand{\doi}
  {\begingroup\let\do\@makeother\dospecials
  \catcode`\{=1 \catcode`\}=2 \doi@aux}
\providecommand{\doi@aux}[1]{\endgroup\texttt{#1}}
\makeatother
\providecommand*\mcitethebibliography{\thebibliography}
\csname @ifundefined\endcsname{endmcitethebibliography}
  {\let\endmcitethebibliography\endthebibliography}{}

\end{document}